\documentclass[oldversion]{aa}
\def\newface{}
\newif\ifpdf\ifx\pdfoutput\undefined\pdffalse\else\pdfoutput=1\pdftrue\fi

\usepackage{graphicx}
\usepackage{natbib}
\usepackage{amsmath}
\bibpunct{(}{)}{;}{a}{}{,}
\usepackage{txfonts}
\def\tens#1{\ensuremath{\mathsf{#1}}}

\newcommand{\A}           {{\tens{A}}}
\newcommand{\F}           {{\tens{F}}}
\newcommand{\FT}          {{\tens{F^\dag}}}

\def\OP#1#2#3#4{{\tens{#1^{#3}_{#2}}#4}}
\def\JS#1#2#3{{\tens{J^{#2}_{#1}}#3}}
\def\MS#1#2#3{{\tens{M^{#2}_{#1}}#3}}
\def\E#1#2#3{{\tens{E^{#2}_{#1}}#3}}

\begin{document}
\title{Correcting direction-dependent gains in the deconvolution of
radio interferometric images}
\titlerunning{Correction of direction-dependent gains during image deconvolution}
%\authorrunning{Bhatnagar, Cornwell, Golap \& Uson}
\author{S. Bhatnagar \inst{1} \and T.J. Cornwell \inst{2} \and
K. Golap \inst{1} \and Juan M. Uson \inst{3}}

\offprints{S. Bhatnagar}

\institute{National Radio Astronomy Observatory
  \thanks{Associated
    Universities Inc. operates the National Radio Astronomy Observatory
    under cooperative agreement with the National Science Foundation},
  1003 Lopezville Road, Socorro, NM, 87801, USA.\\
  \email{sbhatnag@nrao.edu} \email{kgolap@nrao.edu}
  \and
  Australia Telescope National Facility, Epping, New South Wales,
  Australia, 2120.\\
  \email{Tim.Cornwell@csiro.au}
  \and
  National Radio Astronomy Observatory$^*$, 520 Edgemont Road,
   Charlottesville, VA, 22903, USA.\\ 
  \email{juson@nrao.edu}
}

\date{Received: 19 December 2007 / Accepted: }

\abstract{Astronomical imaging using aperture synthesis telescopes
  requires deconvolution of the point spread function as well as
  calibration of instrumental and atmospheric effects.  In general,
  such effects are time-variable and vary across the field of view as
  well, resulting in direction-dependent (DD), time-varying gains.
  Most existing imaging and calibration algorithms assume that the
  corruptions are direction independent, preventing even moderate
  dynamic range full-beam, full-Stokes imaging.  We present a general
  framework for imaging algorithms which incorporate DD errors. We
  describe as well an iterative deconvolution algorithm that corrects
  {\em known} DD errors due to the antenna power patterns (including
  errors due to the antenna polarization response) as well as pointing
  errors for high dynamic range full-beam polarimetric imaging.  Using
  simulations we demonstrate that errors due to realistic primary
  beams as well as antenna pointing errors will limit the dynamic
  range of upcoming higher sensitivity instruments like the EVLA and
  ALMA and that our new algorithm can be used to correct for such
  errors.  We show that the technique described here corrects for
  effects that can be described as approximate unitary operators in
  the interferometric measurement equation, such as those due to
  antenna pointing errors and non-azimuthally symmetric antenna power
  patterns.  We have applied this algorithm to VLA 1.4~GHz
  observations of a field that contains two ``4C'' sources and have
  obtained Stokes~I and~V images with systematic errors that are one
  order of magnitude lower than those obtained with conventional
  imaging tools. Residual systematic errors that are seen at a level
  slightly above that of the thermal noise are likely due to
  selfcalibration instabilities that are triggered by a combination of
  unknown pointing errors and errors in our assumption of the shape of
  the primary beam of each antenna.  We hope to present a more refined
  algorithm to deal with the fully general case in due course.  Our
  simulations show that on data with no other calibration errors, the
  algorithm corrects pointing errors as well as errors due to known
  asymmetries in the antenna pattern.}{}{}{}{}

\maketitle

%%%%%%%%%%%%%%%%%%%%%%%%%
\section{Introduction}
%%%%%%%%%%%%%%%%%%%%%%%%%

Astronomical observations made with interferometric radio telescopes
suffer from variable gain effects which can be broadly classified as
direction-independent (DI) and direction-dependent (DD) errors. The
complex instrumental gains due to the electronic devices that follow
the feed elements are directionally independent, whereas the
time-varying gains due to the antenna primary beams provide an example
of direction-dependent gains. Direction-independent effects may be
corrected separately from imaging, so most processing algorithms have
been limited to treating such effects. Direction-dependent gains are
more difficult to incorporate since they must be applied during
imaging which has slowed progress thus far. This must now change,
however, because direction dependent gains are expected to limit
observations with existing as well as next generation telescopes
presently under construction. Indeed, a number of deep observations
made with present telescopes have been limited already by
direction-dependent, time-variable errors.  In this, the first of two
papers, we present an algorithmic framework that allows incorporation
of a class of directionally dependent gain effects during
deconvolution. A key part of this framework is the provision of an
efficient transform between the data and image domains. As an example
of the application of this framework, we demonstrate that the effects
of known antenna pointing errors and beam polarization can be
corrected during imaging.  We also discuss error propagation and the
required computing resources. A second paper will describe an
algorithm constructed within this framework that allows solving for
parameters that describe such gain changes.

%%%%%%%%%%%%%%%%%%%%%%%%%%%%%%%%%%%
\section{The Measurement Equation}
%%%%%%%%%%%%%%%%%%%%%%%%%%%%%%%%%%%

The measurement equation that describes astronomical imaging using
aperture synthesis telescopes can be compactly written using the
Hamaker-Bregman-Sault notation \citep{HBS1} as\footnote{\newface All
expressions in this paper are in the signal-domain polarization frame
(the linear or circular polarization bases).  Conversion to and from
the Stokes frame can be done by the application of an appropriate
co-ordinate transform operator~\citep{HBS1}.  Note that only
the minor cycle of the deconvolution iterations operate in the Stokes
frame (see section~\ref{SEC_DIRDEPCORR}).}:
\begin{equation}
\label{ME}
V^{Obs}_{ij} = \MS{ij}{}{}\int \MS{ij}{Sky}{}(\vec{s}) I(\vec{s})
e^{2\pi\iota \vec{s}\cdot\vec{b_{ij}}} d\vec{s}
\end{equation}
where $V^{Obs}_{ij}$ is the observed full polarization visibility
vector, $\MS{ij}{}{}$ and $\MS{ij}{Sky}{}(\vec{s})$ are the Mueller
Matrices \citep{Mueller} for {\newface DI and DD} gains respectively,
$\vec{s}$ is a direction in the sky, $I$ is the image and
$\vec{b_{ij}}$ is the vector that describes the projected separation
between the antennas $i$ and $j$ in units of the wavelength of the
observation.

Measurements sample this equation, providing constraints on the
unknowns on the right hand side. The sky brightness distribution
$I(\vec{s})$ has to be estimated in the presence of known or unknown gain
terms $\MS{ij}{}{}$ and $\MS{ij}{Sky}{}$. However, the measurement equation
cannot simply be inverted as it is not a Fourier
transform. Furthermore, as is the case for simpler forms of the
measurement equation, the visibility is sampled only at a limited set
of points so there is insufficient information to determine the
solution exactly. We follow the normal terminology and call the
estimation of the sky brightness ``deconvolution'' though it is only
vaguely related to the typical deconvolution of a shift-invariant
point-spread function \citep{Andrews_Hunt}. The other part of imaging is
the (necessary) estimation and correction of the gains either assuming
that the sky brightness distribution is known (``calibration'') or determined
simultaneously (``self-calibration'').

Most often, deconvolution and self-calibration are performed using simple
steepest-descent algorithms. The residuals are minimized in the least
square sense by minimizing\ $\chi^2$ which can be expressed in
terms of the data ($\vec{V}$) and the model data ($\vec{V^M}$) as 
\begin{equation}
\chi^2 = \left[\vec{V}-\vec{V}^M\right]^{\dag} \Lambda \left[\vec{V} - \vec{V}^M\right]
\end{equation}
{\newface where $\Lambda$ is the inverse of the measurement noise
covariance matrix.  The sky-brightness model, the unknown gain terms
as well as suitable imaging weights are included in $\vec{V^M}$.
Gradients of $\chi^2$ with respect to the sky brightness model and the
gain terms may be calculated straightforwardly and used in an
iterative minimization algorithm to solve for the sky brightness model
and unknown gains
\citep{Schwarz_Clean,Asp_Clean}.}  Often, this results in complex and
non-linear algorithms but these seem to work well most of the time. To
make a practical algorithm, it is necessary to find efficient ways of
calculating the gradients of $\chi^2$ with respect to the unknowns. In
this paper, we consider the case of estimating the sky brightness in
the presence of {\em known} directionally dependent gain terms. In a
subsequent paper, we will consider the estimation of parameters
describing {\em unknown} gain terms.

There are cases in which the deconvolution and correction for the
Mueller matrix can be decoupled. For example, direction dependent
effects which are identical for all the measurements can be removed by
dividing the deconvolved image by {\newface $\MS{}{Sky}{}$}. The correction
for an azimuthally symmetric and time-constant antenna power pattern
provides one such example. The deconvolution is then performed on the
entire data set, ignoring the antenna power pattern, whose inverse
function is applied to the image only after the deconvolution and
self-calibration have been completed. However, this assumption often
breaks down. {\newface For example, the antenna power pattern for most
practical antenna geometries is azimuthally asymmetric
(because of off-axis feed position, asymmetric subreflector, feed-legs, \ldots)
and rotates on
the sky for azimuth-elevation mount telescopes resulting in off-axis
gains which vary with parallactic angle (PA).}  In such case,
processing time slices of the data independently might lower the range
of the gain variations in each subset. The final deconvolved images
for each subset are averaged post-deconvolution. This is
straightforward and often used, but can be expected to be sub-optimal
because the deconvolution step is inherently non-linear and higher PSF
sidelobes for individual subsets increase the level of
(non-symmetric) deconvolution errors in each sub-image. Hence it would
seem preferable to follow a procedure that applied the corrections
while imaging the full data set.

%%%%%%%%%%%%%%%%%%%%%%%%%
\section{Example of directionally dependent gains}
%%%%%%%%%%%%%%%%%%%%%%%%%
In this section, we consider in detail an example of directionally
dependent gains - the antenna far-field voltage pattern.

The far-field voltage pattern is the Fourier transform of the antenna
illumination function \citep{KRAUS}.  Thus it is typically the case
that because of the details of the antenna geometry (such as quadrupod
legs) and feed design, the antenna voltage patterns are azimuthally
asymmetric.  Furthermore, the polarization response of the antenna
will vary away from the antenna optical axis due to antenna geometry
and the physics of the reflection of electromagnetic waves from curved
surfaces.  In addition, as an interferometric array composed of
altitude-elevation mounted antennas tracks a region of the sky, these
asymmetrical antenna voltage patterns rotate on the sky. This, along
with significant time varying antenna pointing errors, makes
{\newface $\MS{}{Sky}{}$} time varying and different for each antenna pair
(interferometric baseline).  Even equatorially mounted antennas share
in this problem to the extent that changes in elevation (temperature)
might deform the antennas due to gravity (dilation).

%%%%%%%%%%%%%%%%%%%%%%%
\subsection{The Jones and Mueller matrices}
%%%%%%%%%%%%%%%%%%%%%%%
The Mueller matrix is an outer product of the two antenna based Jones
matrices \citep{Jones,HBS1}.  A full direction-dependent polarimetric
description requires a Jones matrix per pixel in the image.  For the
two orthogonal polarizations, labeled $p$ and $q$, the Sky Jones
matrix as a function of direction is given by:
\begin{equation}
\label{JONES}
\JS{i}{Sky}{(\vec{s})}=
\left[
\begin{array}{cc}
\JS{i}{p}{} & -\JS{i}{pq}{}\\
\JS{i}{qp}{} & \JS{i}{q}{}\\
\end{array}
\right]
\end{equation}
The super-scripts $pq$ and $qp$ represent leakage of the
$q$-polarization signal to $p$-polarization signal and vice-versa.
The diagonal elements correspond to the antenna voltage patterns on
the sky while the off-diagonal elements correspond to the polarization
leakage terms ($p\rightarrow q$ and $q\rightarrow p$) due to
instrumental leakage (antenna geometry, electronics) and/or
atmospheric, ionospheric or other transmission effects such as Faraday rotation.

The full direction-dependent Sky Mueller matrix $\MS{ij}{Sky}{}$ for baseline 
$i$-$j$ is a $4\times 4$ matrix:
\begin{eqnarray}
\label{MUELLER}
\MS{ij}{Sky}{(\vec{s})} &=& 
\JS{i}{Sky}{(\vec{s})}\otimes\JS{j}{Sky^{\textstyle*}}{(\vec{s})}= \nonumber\\
&&\left[
\begin{array}{cccc}
\JS{i}{p}{}\JS{j}{p^{\textstyle *}}{}  & -\JS{i}{p}{}\JS{j}{pq^{\textstyle *}}{} & -\JS{i}{pq}{}\JS{j}{p^{\textstyle *}}{} &  \JS{i}{pq}{}\JS{j}{pq^{\textstyle *}}{}\\
\JS{i}{p}{}\JS{j}{qp^{\textstyle *}}{} &  \JS{i}{p}{}\JS{j}{q^{\textstyle *}}{}  & -\JS{i}{pq}{}\JS{j}{qp^{\textstyle *}}{}& -\JS{i}{pq}{}\JS{j}{q^{\textstyle *}}{}\\
\JS{i}{qp}{}\JS{j}{p^{\textstyle *}}{} & -\JS{i}{qp}{}\JS{j}{pq^{\textstyle *}}{}&  \JS{i}{q}{}\JS{j}{p^{\textstyle *}}{}  & -\JS{i}{q}{}\JS{j}{pq^{\textstyle *}}{}\\
\JS{i}{qp}{}\JS{j}{qp^{\textstyle *}}{}&  \JS{i}{qp}{}\JS{j}{q^{\textstyle *}}{} &  \JS{i}{q}{}\JS{j}{qp^{\textstyle *}}{} &  \JS{i}{q}{}\JS{j}{q^{\textstyle *}}{}\\
\end{array}
\right]
\end{eqnarray}
\begin{figure*}[ht!]
%\vskip -3.5in
%\hskip -0.25in
\centering
%\hbox{
%\includegraphics[width=7cm]{Figures/Plane1}
%\includegraphics[width=7cm]{Figures/Plane4}
\includegraphics[width=7cm]{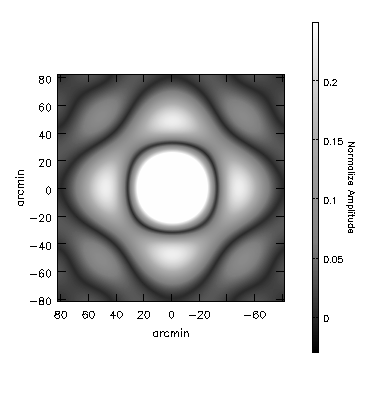}
\includegraphics[width=7cm]{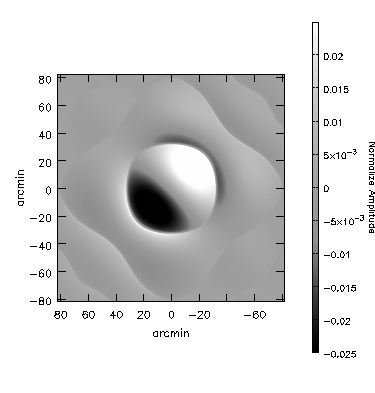}
%}
\caption{\small The image in the left panel shows a typical (first)
 diagonal term for the VLA at 1.4~GHz ($\JS{i}{R}{}\JS{j}{R^{\textstyle *}}{}$) of
 the Sky Mueller matrix.  The pattern is off-center due to the
 off-axis location of the feeds.  The image in the right panel shows
 the difference between the first and second diagonal term
 ($\JS{i}{R}{}\JS{j}{R^{\textstyle *}}{}-\JS{i}{R}{}\JS{j}{L^{\textstyle *}}{}$).
 $\JS{}{Sky}{}$ was evaluated at PA=0 and normalized by the peak of
 one of the diagonal terms (the peak values of both diagonal terms are the
 same).}
\label{JONES_DIAG}
\end{figure*}
The diagonal elements of this matrix are the antenna power patterns
for the four polarization products, whereas the off-diagonal products
incorporate the cross-polarization leakage terms.  For the VLA
antennas, where the two circular polarization power
patterns are squinted with respect to each other, the power pattern
for the parallel hand ($\JS{i}{R}{}\JS{j}{R^{\textstyle *}}{}$) and the difference
between a parallel hand and a cross hand product
($\JS{i}{R}{}\JS{j}{R^{\textstyle *}}{} - \JS{i}{R}{}\JS{j}{L^{\textstyle *}}{}$) are shown in
Fig.~\ref{JONES_DIAG} (super-script $\tens{R}$ and $\tens{L}$ denotes
the right- and left-circular polarizations respectively).  The main
lobe of the power pattern is azimuthally asymmetric and, clearly,
highly asymmetric in the first sidelobe.  This asymmetry is due to
aperture blockage by the feed and the feed-legs.  The two parallel-hand
power patterns ($\JS{i}{R}{}\JS{j}{R^{\textstyle *}}{}$ and
$\JS{i}{L}{}\JS{j}{L^{\textstyle *}}{}$ diagonal terms) are also not identical
because of differences between the power patterns for the two
orthogonal polarizations. Rotation of these patterns on the sky as a
function of PA leads to time varying, direction-dependent gains.  Even
differences between two diagonal terms, for example
$\JS{i}{R}{}\JS{j}{R^{\textstyle *}}{} - \JS{i}{R}{}\JS{j}{L^{\textstyle *}}{}$, are on the order
of a few percent and vary with position within the beam.  In
addition, differences between antennas (focus, surface accuracy,
pointing) will lead to second-order differences in the values of any
given term.  Therefore, an assumption of diagonal form for
$\MS{ij}{Sky}{}$ will lead to errors in the image domain - particularly
in the presence of
strong sources located in the outer parts of the main lobe as
well as in the first few sidelobes.
\begin{figure*}[ht!]
%\vskip -3.5in
%\hskip -0.25in
\centering
%\hbox{
\includegraphics[width=7cm]{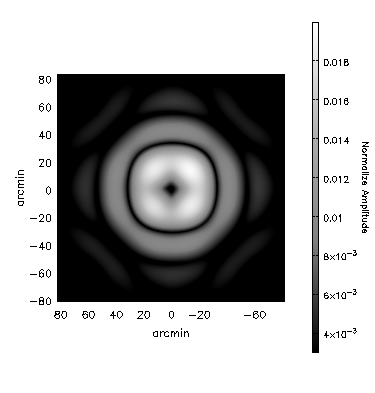}
\includegraphics[width=7cm]{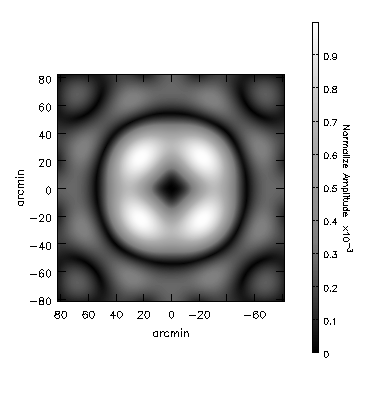}
%}
\caption{\small The off-diagonal terms of the Sky Mueller matrix for
  VLA antennas at 1.4~GHz.  The image in the left panel is of first
  order in the antenna leakage ($\JS{i}{R}{}\JS{j}{RL^{\textstyle *}}{}$).  The
  image in the right panel is the second order term in antenna leakage
  ($\JS{i}{RL}{}\JS{j}{RL^{\textstyle *}}{}$). $\JS{}{Sky}{}$ was evaluated at PA=0
  and normalized by the peak of one of the diagonal terms (the peak values of
  both diagonal terms are the same).}
\label{JONES_OFFDIAG}
\end{figure*}

For high dynamic range imaging ($\ge$few$\times 1000$), the
off-diagonal terms of the Mueller matrix are non-negligible, vary
across the entire beam and, typically, increase substantially with
distance from the center. Figure~\ref{JONES_OFFDIAG} shows the %check
figure is the correct one typical off-diagonal terms for the VLA (the
cross polar power patterns $\JS{i}{R}{}\JS{j}{RL^{\textstyle *}}{}$
and $\JS{i}{RL}{}\JS{j}{RL}{}^{\textstyle *}$).  These are the higher
order leakage terms and are purely due to the leakage of the
orthogonal polarization signals into the complementary polarization in
the signal path from various antennas The term shown in the
left-hand-side panel is the first order leakage term and has peak
amplitude relative to the diagonal terms of $\sim10^{-2}$.  Since this
is comparable to the difference between the diagonal terms, ignoring
this term will result in imaging artifacts similar to those due to the
assumption that parallel- and cross-hand power patterns are identical.
The second order term $\JS{i}{RL}{}\JS{j}{RL^{\textstyle *}}{}$, shown
in the right hand side panel, has an amplitude of $\sim10^{-4}$.
Consequently, for high dynamic range imaging ($\ge$few$\times 1000$),
the $\MS{ij}{Sky}{}$ cannot be approximated as even a {\em diagonally
dominant} matrix.  For even moderate dynamic range full-Stokes,
full-beam imaging, this difference needs to be taken into account.

Existing calibration procedures split the errors into two antenna
based Jones matrices - the $\OP{G}{}{}{}$ and $\OP{D}{}{}{}$ matrices
for complex gain and polarization leakage, respectively. $\OP{G}{}{}{}$
is assumed to be purely diagonal while $\OP{D}{}{}{}$ is unity along the
diagonal and the off-diagonal terms are the leakage gains
\citep{HBS1}.  The directional dependence of these terms is ignored and
the values of the complex gains and leakage gains at the center of the
field are used throughout the beam.  Typically, polarization leakage
is small near the optical axis of the antenna.  Hence,
for imaging compact sources at the center of the beam, the Sky
Mueller matrix is diagonally dominant and the above approximation is
justified. When imaging fields with significant emission throughout
the primary beam, this approximation will lead to artifacts and
significantly lower image fidelity away from the image center. This is
even more true for the case of mosaic imaging \citep{Cornwell88}
where there is significant flux density throughout the primary beam for
most pointings.

Therefore for high dynamic range full beam imaging, full treatment of
polarization has to be kept in the entire imaging and calibration
process.  We refer to this as J-Matrix based imaging and
calibration. Stokes images have to be made from the linear addition of
the visibilities from all polarization products, weighted by the
appropriate terms of the Sky Mueller matrix.  Strictly speaking, even
conventional Stokes-I imaging using only the parallel hand visibilities is
incorrect.  For moderate dynamic range full beam, full Stokes imaging,
it may be possible to use only the diagonal terms of the Sky Mueller
matrix.  Note that the computational load of J-Matrix based imaging is
a factor of four higher than the corresponding load
for conventional imaging (where the
J-Matrix is assumed to be purely diagonal).

%%%%%%%%%%%%%%%%%%%%%%%%%
\section{A deconvolution algorithm incorporating direction-dependent
  gain correction}
%%%%%%%%%%%%%%%%%%%%%%%%%
\label{SEC_DIRDEPCORR}

As described above, most iterative deconvolution algorithms derive
updates ($\Delta\vec{I}^M$) to the existing model image ($\vec{I}^M$)
from the gradients of chi-square with respect to the unknown sky
brightness:
\begin{equation}
\Delta\vec{I^M}=-\mathcal{\OP{C}{}{}{}}{\partial\chi^2/\partial\vec{I}^{M}}=\OP{C}{}{}{}\vec{I}^R
\end{equation}
where $\OP{C}{}{}{}$ is a scaling term, either a constant or the
inverse Hessian (or an approximation thereof).

Typically, the model image is iteratively improved as:
\begin{equation}
\vec{I}^M_i =\OP{T}{}{}{}\left(\vec{I}^M_{i-1}, \left[{\vec{I}^R_i}\right]\right)
\end{equation}
where $\vec{I}^R_i$ is the Fourier transform of the residual
visibilities ($\vec{V}^R$) and $\vec{I}^M_i$ is the cumulative
model at the $i^{th}$ iteration. The operator $\OP{T}{}{}{}$ selects
part of the gradient image. For the H\"ogbom Clean algorithm
\citep{Hogbom_Clean}, $\OP{T}{}{}{}$ simply updates the model by adding a
scaled version of the the peak in the image.

\begin{figure*}[ht!]
%\vskip -3.5in
%\hskip -0.25in
%\hbox{
\centering
\includegraphics[width=7cm]{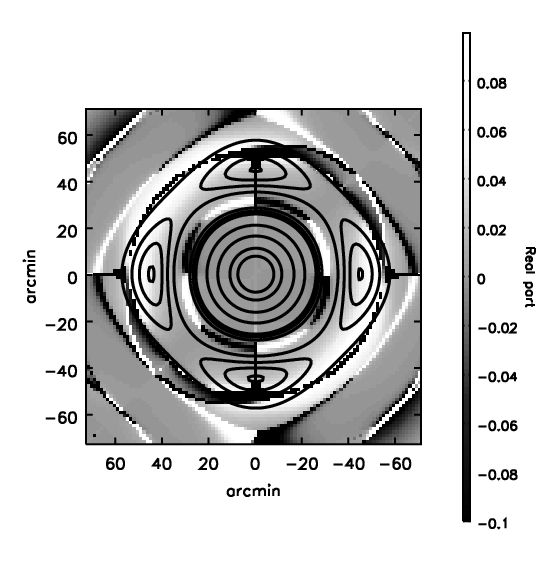}
\includegraphics[width=7cm]{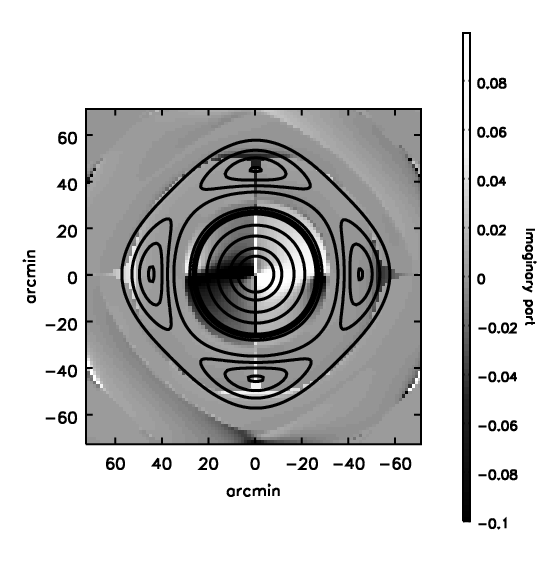}
%}
\caption{\small The off-diagonal term of
  $\JS{i}{Sky^\dag}{}\JS{i}{Sky}{}$. Images in the left and right
  panels show the real and imaginary parts respectively.  The
  overlayed contours correspond to $max \left( \JS{i}{Sky}{}[0]
  \right) \times [0,0.1,0.15,0.18,0.2,0.4,0.6,0.8,0.9]$. The absolute
  maximum value inside the first null is $\sim 0.2$.}
\label{Ji01}
\end{figure*}
Following terminology established by Clark (1980), the calculation of
the derivative for a given estimate is called the {\em major} cycle,
and the application of the $\OP{T}{}{}{}$ operator is called the {\em
minor} cycle. {\newface The minor cycle typically operates in the
Stokes frame.  The operator $\OP{T}{}{}{}$ includes conversions
between the signal-domain polarization frame and the Stokes frame
using an appropriate coordinate transform operator \citep{HBS1}.}

The major cycle can be broken into two calculations:
\begin{description}
\item[forward:] In the forward step, the model visibilities for
  baseline $i$-$j$ are calculated from the existing model image
  $\vec{I^{M}}$, using the equation:
\begin{equation}
\label{ME_M}
\vec{V}^{M}=\A\vec{I}^{M}
\end{equation}
where the operator $\A$ is the measurement matrix.
\item[backward:] In the backward step, the residual visibilities
  ($\vec{V^R}$) are propagated
backwards to the image plane using the equation:
\begin{equation}
\vec{I}^{R}=\left[\A^{\dag}\A\right]^{-1}\A^\dag\vec{V}^{R}
\end{equation}
\end{description}

The forward calculation must be done with high accuracy, but since the
overall approach is iterative, the backward calculation may be
performed with lower accuracy \citep{Schwab1983,Cotton1989} When using
the FFT algorithm for computing the Fourier transform, the gridded
visibilities $\F\vec{I}^M$ are interpolated from a regular grid and
re-sampled at the measured $(u,v,w)$ points as:
\begin{equation}
\label{GRIDDED_ME}
\vec{V}^{M^\prime}(u_{ij},v_{ij},w_{ij}) = \left(\OP{G}{}{}{}\left[\F 
\vec{I}^M\right]^g\right)(u_{ij},v_{ij},w_{ij})
\end{equation}
where $\OP{G}{}{}{}$ is the interpolation operator, $\F$ is the
Fourier transform operator and the superscript $g$ indicates data on a
regular grid.  The backward calculation will correspond to the
application of $\left[\OP{G}{}{}{}\F\right]^\dag$. The operator
$\OP{F}{}{}{}$ is unitary.  If $\OP{G}{}{}{}$ is at least
approximately unitary, $\OP{G}{}{}{}^\dag$ can be used as the
interpolation operator for re-sampling the data on a regular grid to
correct for the effects of $\OP{G}{}{}{}$ in the image.  Applying
$\OP{G}{}{}{}$ and $\OP{G}{}{}{}^\dag$ as part of the re-sampling
operations require that both of these operators have finite support.  In
principle, any operator for which the inverse exists can be used,
provided the inverse operator also has a finite support.  However,
since the inverse of an operator $\OP{X}{}{}{}$ involves
$det\left(\OP{X}{}{}{}\right)^{-1}$, it is difficult to imagine
$\OP{X}{}{}{}^{-1}$ with a support size comparable to $\OP{X}{}{}{}$.
An approximate inverse operator with finite support for our case can
be constructed by using $\OP{G}{}{}{}^\dag$ for re-sampling the data
(left-hand side of Eq.~\ref{GRIDDED_ME}) and then dividing the
resulting image by $det\left(\OP{FG}{}{}{}\right)$.

The similarity between Eqs.~\ref{GRIDDED_ME} and \ref{ME_M} indicates
that direction-dependent gains can be incorporated as part
of the deconvolution iterations by using an efficient algorithm for
the forward and backward calculations. We have chosen to use a
technique similar to that used in the {\tt w-projection} algorithm to
correct for the effects of non co-planar baselines
\citep{W_PROJECTION}.  As discussed in section~\ref{SEC_JONES},
an approximately unitary operator $\E{ij}{}{}$ can be
constructed as the Fourier transform of Eq.~\ref{MUELLER}.  For our
purpose, using $\E{ij}{^\dag}{}$ as the interpolation operator for
gridding the visibilities on a regular grid and using FFT to invert
the gridded visibilities would suffice. The accuracy of the forward
calculation is proportional to the accuracy of $\E{ij}{}{}$ which can,
in principle, be arbitrarily precise (e.g. by accurate measurement of the
antenna voltage pattern).  An iterative deconvolution scheme using
such transforms should ultimately drive the residual image to be
noise-like, although it would seem desirable to limit the number of free
parameters introduced in the process.
Note that since the final model image is iteratively
built using accurate computations only in one direction, the
intermediate residual dirty images have no physical meaning as is
usually the case. 

\subsection{Structure of the Sky Jones matrix}

\label{SEC_JONES}
$\JS{}{Sky}{}$ affects the measurements as described by Eq.~\ref{ME}.
The forward and backward transforms discussed above crucially depend
on $\E{ij}{}{}$ being at least approximately unitary.  Since
$\MS{ij}{Sky}{}$ is an outer product of antenna based Jones matrices
(Eq.~\ref{JONES}) and $\E{ij}{}{} =
\mathcal{FT}(\MS{ij}{Sky}{})$ where $\mathcal{FT}$ represents the
element-by-element Fourier transform of its argument, for our
purpose it is sufficient to ensure that the $\JS{i}{Sky}{}$ is
approximately unitary\footnote{It can be shown that if $\JS{i}{Sky}{}$
is unitary, then so is $\JS{i}{Sky^\dag}{}\otimes\JS{i}{Sky}{}$}.  The
diagonal terms of $\JS{i}{Sky^\dag}{}\JS{i}{Sky}{}$ (of the form
$\JS{i}{p}{}\JS{i}{p^{\textstyle *}}+\JS{i}{pq^{\textstyle
*}}{}\JS{i}{pq}{}$) correspond to the ideal (un-squinted) power
patterns and are nearly equal to each other.  Fig.~\ref{Ji01} shows
the real and imaginary parts of the off-diagonal term normalized by
$det \left(\JS{i}{Sky}{}\right)$. The peak amplitude is about two
orders of magnitude lower than the diagonal term making $\JS{i}{}{}$
approximately unitary.  Image plane corrections therefore can be
incorporated as part of the image deconvolution procedure by using
$\E{ij}{}{}=\mathcal{FT}\left[\MS{ij}{Sky}{}\right]$ and
$\E{ij}{\dag}{}$ as part of the forward and reverse transforms between
the visibility and image domains for baseline $i$-$j$.

\subsection{The overall deconvolution algorithm}

Our deconvolution algorithm proceeds as follows:

\begin{enumerate}
\item Initialize: Set the initial model image to zero or to a model
  using apriori knowledge of the sky emission (for example a model obtained with conventional techniques).
\item Major cycle:
\label{PAINC}
\begin{itemize}
\item Forward: Compute the residual visibilities
  $\vec{V^{Obs}}-\vec{V^{M}}$ using the observed visibilities
  $\vec{V^{Obs}}$ for each polarization product.
\item Backward: Compute the residual image using Eqs.~\ref{GRIDDING}
  and \ref{IMAGING} below.
\end{itemize}
\item Minor cycle: Update the model image applying some operator
  $\OP{T}{}{}{}$.
\item Go to~\ref{PAINC} until convergence is achieved, typically quantified by suitable
stopping criteria (noise level, distribution of residuals, etc.).
\item Smooth the deconvolved image by the resolution element and add
  back the residuals.
\end{enumerate}

%%%%%%%%%%%%%%%%%%%%%%%%%
\section{Antenna polarization and pointing error correction}
%%%%%%%%%%%%%%%%%%%%%%%%%
In the following section, we describe how the forward and backward
calculations can account correctly for polarization leakage and pointing
errors.

\begin{figure*}[ht!]
\centering
\vbox{
\hskip 0.3cm
\hbox{
\includegraphics[width=8cm]{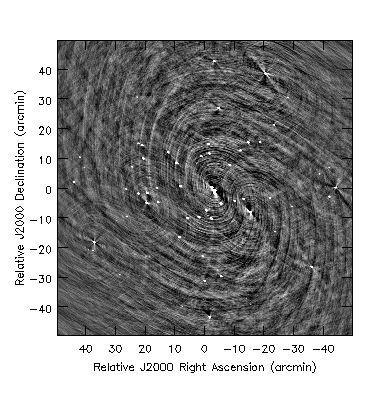}
\includegraphics[width=8cm]{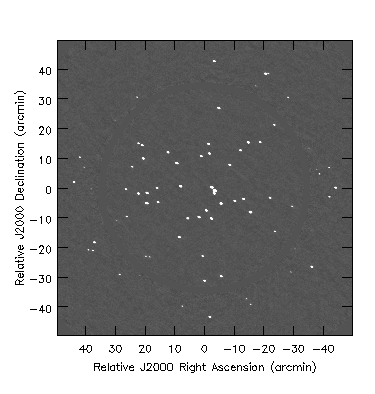}
}
\hskip 0.3cm
\hbox{
\includegraphics[width=8cm]{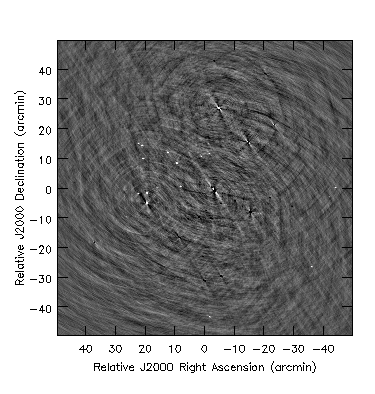}
\includegraphics[width=8cm]{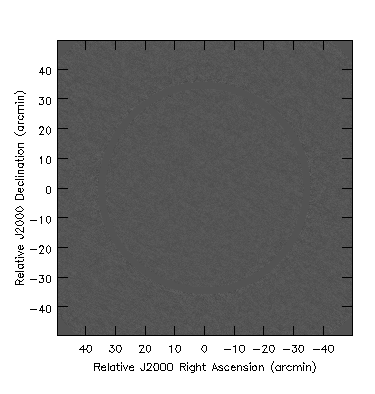}
}
}
\caption{\small The top row shows the Stokes-I images and the bottom
  row shows the Stokes-V images.  The images on the left were made
  without squint and pointing correction while those on the right had
  both corrections applied.  The deconvolution errors seen around the
  strongest sources are due to the antenna pointing errors and time
  varying direction dependent gain due to the rotation of azimuthally
  asymmetric antenna power patterns. These images were made using a
  linear transfer function with the gray scales in the range
  $-20~\mu$Jy/beam (black) and $+40~\mu$Jy/beam (white).  The RMS
  noise in the off-source regions of the images in the left and the
  right panels is $10~\mu$Jy/beam and $1~\mu$Jy/beam respectively.}
\label{EX1}
\end{figure*}

\subsection{Forward calculation}
\label{SEC_FORWARD}
In the absence of antenna pointing errors, the operator $\E{ij}{P}{}$
is the auto-correlation of the ideal antenna illumination patterns for
polarization product $\tens{P}$.  It has finite support and is
also approximately unitary.  Therefore, it has the required properties
to be used to realize the transforms necessary in a deconvolution
algorithm to correct for primary beam effects.  In the presence of
antenna pointing errors, the operator $\E{ij}{P}{}$ is different for
each baseline $i$-$j$.  For small pointing errors compared to the half
power beam width, pointing errors contribute a linear phase gradient
across the aperture.  Therefore, the full $\E{ij}{P}{}$ including antenna
pointing errors can be efficiently evaluated by separating
it into terms that include the effects which are equal for all
antennas (e.g.  the polarization squint of the VLA antennas) and effects
which vary between antennas (e.g. the antenna-based pointing offsets)
as:
\begin{equation}
\E{ij}{P}{} = \E{}{P^\circ}{} f(\phi_i - \phi_j) e^{\iota (\phi_i+\phi_j)}
\label{EIJ}
\end{equation}
where $\phi_i$ is the pointing offset for antenna $i$ and
$\E{}{P^\circ}{}$ is the auto-correlation of the ideal antenna
illumination pattern.  The function $f$ represents the de-correlation
that the signal suffers at each baseline due to the antenna pointing
errors.  This function is unity at the origin ($f(0)=1.0$) and for
voltage patterns with finite support, it will be a monotonically
decreasing function of its argument (for typical illumination
functions) \footnote{This might not be a monotonically decreasing
  function for all ways in which a feed might illuminate a
  secondary/primary reflector.}.  However, its exact form will depend
upon the actual form of the voltage patterns.  For small pointing
errors (few percent of the half-power beam width), it will be close to
unity to the first order.  When using $\E{ij}{P}{}$ as the visibility
plane filter for baseline $i$-$j$, with no pointing errors the
predicted visibilities will correspond to a sky tapered by the
corresponding power pattern (as it should be).  With pointing errors,
the predicted visibilities will include the effects of pointing
errors.

\subsection{Backward calculation}

The backward calculation can be realized by using $\E{ij}{P^\dag}{}$
as the interpolation operator for re-sampling 
$\vec{V}^\tens{P}(u_{ij},v_{ij})$ (the visibilities for
the polarization product $\tens{P}$) on a regular grid at pixels labeled
by indices $(n,m)$ as:
\begin{equation}
\vec{V}^{\tens{P},G}(n\Delta u,m\Delta v) = 
\left(\E{ij}{P^\dag}{}{\vec{V}^\tens{P}}(u_{ij},v_{ij})\right)(n\Delta 
u,m\Delta v)
\label{GRIDDING}
\end{equation}
where the superscript $\tens{G}$ is used to indicate a regular grid.  The images corresponding to the gridded visibilities are then computed
as
\begin{equation}
\vec{I}^d = det\left(\overline{\FT \left[\E{}{P^\dag}{}\right]}\right)^{-1}\FT \vec{V}^{\tens{P},G}
\label{IMAGING}
\end{equation}
The net effect of $\E{ij}{}{}$ in the image domain (expressed by
$\overline{\FT
\left[\E{}{P^\dag}{}\right]}$ in Eq.~\ref{IMAGING} above) is
therefore averaged over all antennas for the entire range of
parallactic angle coverage.
% Make sure that no intended meaning has been lost in the rewording!

\begin{figure*}[ht!]
\centering
\includegraphics[width=8cm]{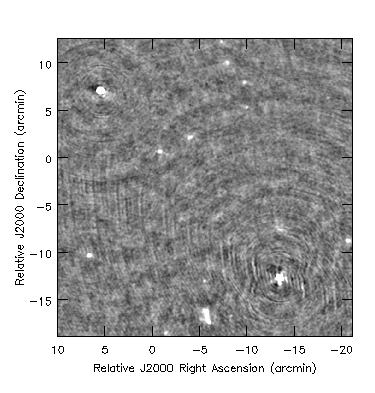}
\includegraphics[width=8cm]{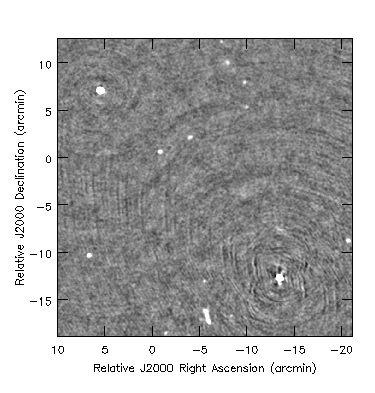}
\caption{\small The Stokes-I images at 1.4~GHz with VLA C-array
  observation.  The left panel shows the deconvolved image without
  corrections of the antenna power pattern variations as a function of
  parallactic angle.  The right panel shows the result from the
  algorithm described in this paper. The two dominant sources, on
  either side of the pointing center have flux densities
  of $\sim854$~mJy/beam and $\sim145$~mJy/beam and the RMS noise is
  $0.15$~mJy/beam.  A linear transfer function with a range $\pm
  1.5$~mJy/beam was used to make these images.
}
  
\label{IC2233_STOKES_I}
\end{figure*}
\begin{figure*}[ht!]
\centering
\includegraphics[width=8cm]{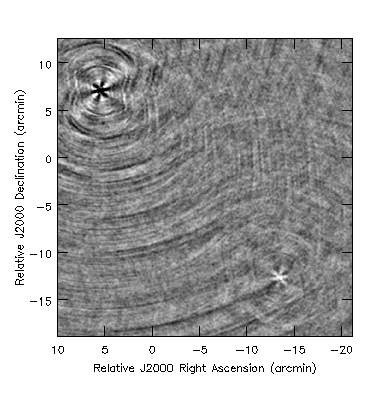}
\includegraphics[width=8cm]{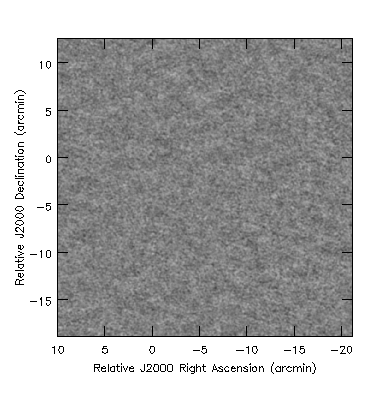}
\caption{\small The Stokes-V images for a 1.4~GHz VLA C-array data.
  The left and right panels show the images without and with
  PB-corrections using a linear transfer function with a range $\pm
  1.5$~mJy/beam was used to make these images. Errors due to the
  polarization squint are greater than thermal noise in the left
  image. The RMS in the right-hand-side image is $\sim0.15$~mJy/beam.
  The negative and positive peaks in this image are
  $\pm0.5$~mJy/beam.  
}
\label{IC2233_STOKES_V}
\end{figure*}

%%%%%%%%%%%%%%%%%%%%%%%%%
\section{Results}
%%%%%%%%%%%%%%%%%%%%%%%%%
\label{SEC_RESULTS}

\subsection{Simulations}
The algorithm was tested for VLA squint and gain variations due to the
rotation of azimuthally asymmetric antenna power patterns on the sky.

\begin{table}
\caption{Simulation parameters}
\label{SIMPARS}
\centering
\begin{tabular}{l c}
\hline\hline
Range of Parallactic Angle($^\circ$) & $183-45$ \\
Number of visibilities & $10^6$ \\
Integration time(sec) & 10 \\
Number of frequency channels & 1 \\
Frequency(GHz)  & 1.4\\
Noise per sample(mJy) & ~1 \\
Max. baseline(Km) & ~3\\
Number of antennas& 27\\
\hline\hline
\end{tabular}
\end{table}

The visibilities were simulated using the $\mathcal{CASA}$\footnote{{\tt
http://casa.nrao.edu}} package with the parameters listed in
Table~\ref{SIMPARS}.  A model for typical sky emission at 1420~MHz was
generated using the NVSS source list.  The actual rendition has 74
point sources with flux densities ranging between 195~mJy and 2~mJy.
A PA increment of $10^\circ$ was used in order to simulate the
rotation of the R- and L-beams on the sky.  The visibilities were
simulated for VLA C-array and an RMS noise of $\sim1~$mJy per
visibility sample was added to simulate an image plane RMS noise of
$\sim1~\mu$Jy/beam.  Without squint correction, the peak and RMS noise
in the resultant Stokes-V image were $\sim2$~mJy and
$\sim10~\mu$Jy/beam respectively.  The Stokes-V image generated using
the algorithm described in this paper was noise-like with an RMS $\sim
1~\mu$Jy/beam.  The squint correction results in an improvement of the
noise figure by a factor of $\sim10$.  The visibilities with pointing
errors were simulated by predicting the model visibilities using the
forward transform (section~\ref{SEC_FORWARD}) with $\E{ij}{}{}$
computed for increments of $10^\circ$ in PA.  A model for the VLA
aperture illumination pattern
\citep{VLA_ILLUMINATION_PATTERN} was used to generate a
non-azimuthally symmetric power pattern.  The model includes the
geometry of the sub-reflector and the feed position as well as the
aperture blockage due to the feed legs and the sub-reflector. The
averages of the pointing errors for each antenna were randomly
distributed between $\pm25\arcsec$ with an RMS of $5\arcsec$.  The
images were deconvolved using standard image deconvolution procedures
and again using the above algorithm.  The Stokes-I images are shown in
Fig.~\ref{EX1}.  As expected, the deconvolution errors were maximal
for the sources around the half-power point and in the first sidelobe
of the power pattern.  These errors were eliminated when the pointing
and squint correction were applied during deconvolution.  The bottom panels
of Fig.~\ref{EX1} show the Stokes-V images without and with pointing
and squint corrections.

%%%%%%%%%%%%%%%%%%%%%%%%%%%%%%%%%%%%
\subsection{VLA 1.4~GHz data}
%%%%%%%%%%%%%%%%%%%%%%%%%%%%%%%%%%%%

The algorithm was also tested for Stokes-I and -V imaging using VLA
1.4~GHz observations of the superthin galaxy IC2233
\citep{Matthews_Uson_08}.  The field contains two strong sources
($\sim854$~mJy/beam and $\sim145$~mJy/beam) on opposite sides of the
pointing center, located at positions of $\sim 75$\% and $\sim 35$\%
primary beam response levels respectively.  The observations were made
in spectral mode with channels of width $\sim 24$~kHz for a total of
$\sim 11.6$~hours in 2~passes with well distributed uv-coverage.  The
line-free channels (11~from the second ``IF pair'') were used for the
tests described here.  The aperture illumination pattern for each
antenna was assumed to be the same and computed using the model for
VLA antennas
\citep{VLA_ILLUMINATION_PATTERN}.  The aperture illuminations were
computed as a function of PA in increments of $1^\circ$.  The expected
thermal noise for this data is $\sim0.13$~mJy/beam. The results of the
imaging run with and without the correction for time-varying primary
beam gains and polarization squint are shown in
Figs.~\ref{IC2233_STOKES_I} and \ref{IC2233_STOKES_V}.  The peak
negative and positive residual in the Stokes-V images without primary
beam correction is $-9.7$~mJy/beam and $2.8$~mJy/beam respectively.
After the primary beam correction, the peaks were $\pm0.5$~mJy/beam
and uncorrelated with location of the bright sources, with an RMS
noise of $0.15$~mJy/beam.

%%%%%%%%%%%%%%%%%%%%%%%%%
\section{Error analysis}
%%%%%%%%%%%%%%%%%%%%%%%%%
\label{SEC_ERRORANALYSIS}
Convergence of the deconvolution iterations is judged by the
statistics in the residual image at convergence.  Errors in the final
residual image purely due to primary beam (PB) effects (within the
main-lobe) can be expressed as:
\begin{equation}
\vec{I^R} = \sum_{\psi} PSF(\psi)\star \left[\Delta PB(\psi) \vec{I^\circ}\right]
\end{equation}
where '$\star$' represents the convolution operator, $\psi$ is the
feed Parallactic Angle, 
$\Delta PB(\psi)$ is the error between the true and the assumed
primary beam model at PA$=\psi$, $\vec{I^\circ}$ is the true sky
distribution and $PSF(\psi)$ is the instantaneous (snapshot) PSF. When
imaging using an azimuthally symmetric PB model, the PB error pattern
is given by $\Delta PB(\psi) = \overline{PB} - PB(\psi)$ where
$\overline{PB}$ is the azimuthally averaged PB.  Rotation of this
error pattern on the sky contributes the dominant systematic errors in
the residual image (and consequently in the final deconvolved image).
The peak residual can be estimated for a point source of flux density
$S$ located at the position of the peak of the error pattern
multiplied by the maximum sidelobe of the instantaneous PSF at
PA=$\psi_i$:
\begin{equation}
\left.I^R\right|_{max} =
\left[\left.PSF_{sidelobe}(\psi_i)\right|_{max}\right] \left[\left.\Delta PB(\psi_i)\right|_{max}\right]S
\end{equation}

\begin{figure}[ht!]
\centering
\includegraphics[width=7cm]{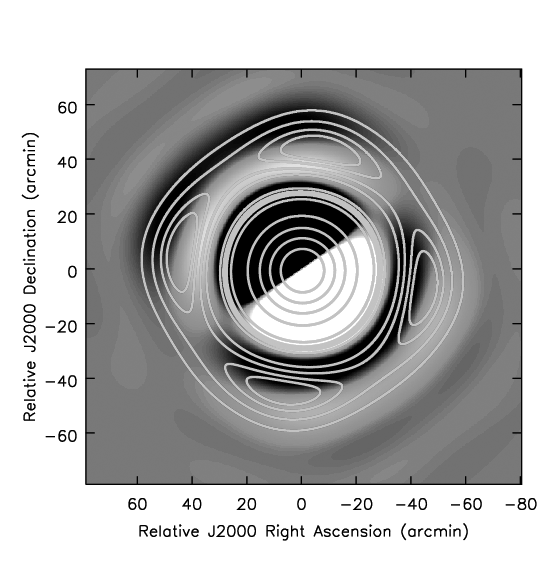}
\caption{\small Model for the VLA 1.4~GHz antenna at
  Parallactic Angle $\sim80^\circ$. The Stokes-V pattern
  ($[PB_{RR}-PB_{LL}]/2$) is shown in colour/gray scale with the
  contours of the Stokes-I power pattern superimposed. Dark regions in
  the gray scale image represent negative values due to the
  polarization squint of VLA antenna.}
\label{PB}
\end{figure}
\begin{figure}[ht!]
\centering
\includegraphics[width=8cm]{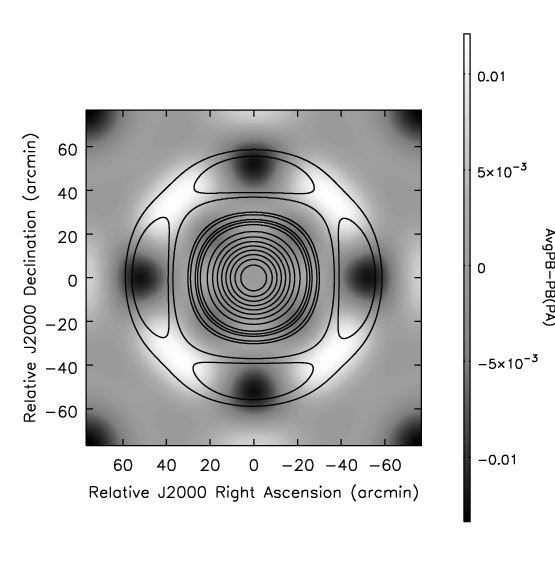}
\caption{\small Difference between an instantaneous Stokes-I PB and an
  azimuthally averaged PB ($\overline{PB}$).  The colour/gray scale image is
  $\Delta PB = PB(\psi_\circ)-\overline{PB}$ while the contours are
  for the $\overline{PB}$.}.
\label{dPB-I}
\end{figure}
\begin{figure}[ht!]
\centering
\includegraphics[width=80mm]{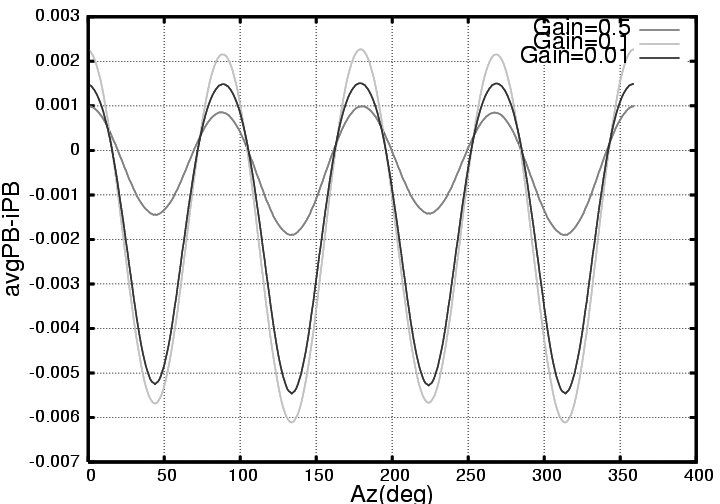}
\caption{\small Azimuthal cut through $\Delta PB$ shown in
  Fig.~\ref{dPB-I} at points where the value of the $\overline{PB}$ is
  at $50\%, 10\%$ and $1\%$ of its peak value.}
\label{dPBCut-I}
\end{figure}

Instantaneous Stokes-I and -V VLA antenna power patterns at 1.4~GHz
are shown in Fig.~\ref{PB}.  The patterns rotate on the sky with PA
which results in time-varying, position-dependent gain across the
field of view.  The $\Delta PB$ and an azimuthal cut through this
error pattern at $50\%$, $10\%$ and $1\%$ point of $\overline{PB}$ are
shown in Figs.~\ref{dPB-I} and ~\ref{dPBCut-I} respectively.  The
contours correspond to $\overline{PB}_{max}\times
[0.01,0.05,0.07,0.2,0.3,0.4,0.5,0.6,0.7,0.8,0.9,0.95]$.  Within the
main-lobe of the Stokes-I beam, the peak of the error pattern is where
$\overline{PB}$ is between $1-10\%$. For $S=1$~Jy,
$\left.PSF_{sidelobe}(\psi_\circ)\right|_{max}=40\%$ (measured for the
test 1.4~GHz, C-array data) and $\left.\Delta
PB(\psi_\circ)\right|_{max}=0.005$, the peak residual will be about
$2$~mJy in Stokes-I.  Peak residuals in Stokes-V would be at the level
of about $10$~mJy.

The deconvolution algorithm described above consists essentially of
approximating the function shown in Fig.~\ref{dPBCut-I} (or
equivalently the 2D function shown in Fig.~\ref{PB}) by a piece-wise
constant function.  The maximum error due to such an approximation can
be estimated using the following equation:
\begin{eqnarray}
I^R_{max}&=&\Delta I^R_{max} \left[\Delta \psi\right]\\
where~\Delta
I^R_{max}&=&S~\left[\left.PSF_{sidelobe}\right|_{max}\right]~\left.\frac{\partial{PB}}{\partial\psi}\right|_{max}\nonumber
\end{eqnarray}
For a required RMS noise in the image of $\eta$, the peak error due to
the piece-wise constant approximation should be 3--5 times smaller
than $\eta$.  The minimum PA increment such that the RMS noise in the
image is not limited by the piece-wise constant approximation would be
given by
\begin{equation}
\Delta \psi \le 3 \eta/\Delta I^R_{max}
\end{equation}
For the 1.4~GHz, VLA C-array test data,
$\left.\frac{\partial{PB}}{\partial\psi}\right|_{max}
\approx~0.0003~deg^{-1}$.  For a PA increment of $10^\circ$, peak
residuals in Stokes-I and -V would be about 1~mJy and 5~mJy
respectively.  With the expected thermal sensitivity of $0.1$~mJy/beam
for the 1.4~GHz test data we used, PA increments of $1^\circ$ were
required.

The PA increment for higher sensitivity telescopes like the EVLA or
SKA such that imaging is not limited by the above approximation will
be much smaller.  This requirement however can be significantly
relaxed by approximating the error function by a piecewise linear
approximation (interpolation of the functions computed at larger PA
increments).  Furthermore, since image interpolation itself can be
expensive, caching of pre-computed aperture functions at appropriate
PA increments will be necessary.  Note that the gridding cost is
relatively insensitive to the number of convolution functions used.  A
hybrid approach of FFT based transforms plus analytical computations
for the strongest sources will probably deliver optimal performance.

%%%%%%%%%%%%%%%%%%%%%%%%%%%%%%%%%%%%
\section{Discussion}
%%%%%%%%%%%%%%%%%%%%%%%%%%%%%%%%%%%%

Some residual deconvolution errors are still left around the second strongest
source in Fig.~\ref{IC2233_STOKES_I}.  The pattern in the residual
image (not shown) suggests that these errors are due to image
pixelation \citep{Pixelation_Errors, Cotton_Uson_2007}.  More sophisticated
parametrization of the sky, independent of the image pixel size
(e.g. as is done in scale sensitive deconvolution algorithms like the
Asp-Clean \citep{Asp_Clean} or MS-Clean) along with the imaging
algorithm described here to correct for DD gains should give better
results.  It is also possible that the residual errors are due to pointing errors
during the observation.  We are investigating this possibility and hope to
report on it in due course.

\subsection{Why not Peeling?}

The algorithm described here corrects for DD gains without loosing the
efficiency advantage of the FFT algorithm.  Our algorithm scales well
in run-time efficiency and implementation complexity for large data
volume, complex field as well as for arrays where antenna elements
cannot be assumed to be identical.

Various variants of the ``Peeling'' algorithm can also be used to
correct for direction-dependent gains.  In this approach antenna based
gains are determined in the direction of each compact source.  These
gains are then used to subtract the contribution of compact sources
from the observed data using a Direct Fourier Transform (DFT) and the
residual visibilities are imaged again. While this is useful in
removing the artifacts due to strong compact sources, since the gains
are determined independently for each direction in the sky, as the
image complexity increases, too many degrees of freedom (DoF) might be
added to the problem.  For crowded fields (large number of compact
sources), this leads to a proliferation of DoFs and potentially to the
problem of over-fitting (the extreme case being when each pixel in the
image has an associated independent gain which gives the best-fit
result).  Since DFTs have to be used to compute residuals, the
computing load is also significantly higher than the corresponding one
for FFT-based computation of residuals.  For complex fields containing
extended emission, this approach quickly becomes numerically un-viable
because of the large number of DoFs included as well as the high
computing and I/O loads involved.  Therefore, while variants of the
Peeling algorithm could have given better results for the particular
1.4 GHz VLA data that we have used in this paper, we did not resort to
Peeling based algorithms.  However since the goal here is to
demonstrate the effectiveness of the algorithm in correcting for
otherwise difficult to correct DD gains, we used this relatively
simple field for our tests so that the advantages and limitations of
our algorithm are brought to the fore.

Of course, using a direct Fourier transform (DFT) for predicting model
visibilities rather than using the FFT algorithm will give the most
accurate results.  While such a brute-force approach might be useful
for simple fields, as mentioned above, the computing cost for even such
simple fields becomes prohibitive for data with more than a few
frequency channels, even when assuming that the various antenna
elements are identical.  For cases where this assumption breaks down,
as it does even for the simple case of random antenna pointing errors,
computing costs are impractically high.  Furthermore, for full-beam,
full-Stokes imaging, which requires use of at least the diagonal terms of
the Mueller matrix (Eq.~\ref{MUELLER}) if not the full matrix, it is
unclear if a brute-force DFT approach will work.

%%%%%%%%%%%%%%%%%%%%%%%%%%%%%%%%%%%%
\subsection{Implications for wideband and mosaic imaging}
%%%%%%%%%%%%%%%%%%%%%%%%%%%%%%%%%%%%
Antenna pointing errors, azimuthally asymmetric aperture
illuminations, wide bandwidths and deconvolution errors due to the use
of discrete pixels for the sky representation all leave residuals that
limit the full-beam imaging dynamic range to $10^4-10^5$.  Therefore,
apart from correcting for the direction dependent effects, for the
highest imaging dynamic range, scale-sensitive decomposition of the
sky might also be necessary \citep{Asp_Clean}. The algorithm
described here can be combined efficiently with scale-sensitive
deconvolution and has the potential of overcoming the above mentioned
imaging dynamic range limit.

The algorithm described here accounts for the time varying gain
variations due to the rotation of the azimuthally asymmetric aperture
illumination with PA.  For imaging with a large bandwidth ratio
(e.g. the ratio of frequencies at the two edges of the observing band
for EVLA will be 2:1), the dominant error term will be the scaling of
the power pattern with frequency.  Sources which will be well within
the main lobe of the primary beam at the lower frequency end of the
band will be outside the main lobe at the higher frequency end (and
may even appear in the first sidelobe).  Since the azimuthal
variations in the power pattern due to feed-leg/sub-reflector blockage
are maximal close to the null and in the first sidelobe, frequency
scaling of the aperture illumination will contribute a first-order
error.

Scaling of the antenna power patterns with frequency in observations
with wide bandwidths can be incorporated in the algorithm described
here by computing the aperture illumination functions at appropriate
increments in frequency.  Alternatively, depending on the required
accuracy, this scaling can be achieved as well by scaling the
co-ordinates with frequency.  Since the computing cost scales weakly
with the number of convolution functions used, the extra computing
load will not be too high.  The cost of computing the aperture
functions itself will be significant, but it is a one-time cost.

Rotation of the sidelobes results in gain variations on the order of a
factor of two in the direction of the sidelobes.  {\newface For mosaic
observations, this will contribute significant time-varying flux
density in most individual pointings; assuming a peak PSF sidelobe of
$10\%$, the error {\em inside} the main lobe of the power pattern will
be at the level of a few percent of the peak flux in the direction of
the first sidelobe of the antenna power pattern.  This will limit the
mosaicking dynamic range significantly, indeed it will be a
first-order effect.  In addition, a second-order effect will be due to
antenna pointing errors.  Correction of both of these effects will be
required for mosaicking instruments presently under construction like
the ALMA and the ASKAP
\citep{ASKAP}.  The general framework and the algorithm described here
can be generalized easily for application to mosaic imaging and could
correct errors due to the rotation of antenna primary beams as well as
pointing errors.  In practice however, the imaging dynamic range might
be limited by the precision with which antenna power patterns and
pointing errors can be determined.}  Furthermore, a similar approach
can be used for imaging with inhomogeneous arrays like CARMA/ALMA
(where not all antennas in the array are identical) as well as with
arrays with multi-feed antennas like the ASKAP.  Finally, for very
high dynamic range imaging with telescopes like LOFAR, SKA, and even
EVLA, nominally identical antenna elements may have variations that
will induce errors higher than the thermal noise limit.  In that
sense, such telescopes will also need to be treated as inhomogeneous
arrays.

%%%%%%%%%%%%%%%%%%%%%%%%%
\section{Conclusions}
%%%%%%%%%%%%%%%%%%%%%%%%%
Existing imaging algorithms ignore the effects of time varying gains
due to antenna pointing errors and rotation of azimuthally asymmetric
antenna power patterns.  As shown in sections~\ref{SEC_RESULTS} and
\ref{SEC_ERRORANALYSIS} using the VLA as an example, residual errors due
to these effects are maximal in the first sidelobe and significant
even within the main lobe of the antenna power pattern.  Simulations
show that the errors due to these effects limit the achievable dynamic
range for sensitive radio interferometers under construction like the
EVLA, ALMA and ASKAP.  The full polarimetric response of the
antenna is also inherently asymmetric due to the physics of reflection
from curved surfaces.  In addition, the voltage patterns of
the two orthogonal polarizations for the VLA antennas are separated on
the sky (polarization squint) resulting in increasing instrumental
Stokes-V as a function of distance from the image center.  Therefore,
for moderate dynamic range full-beam, full-Stokes imaging, the Sky
Jones matrix for the VLA antennas cannot be assumed to be
scaled-identity or even diagonal.  Even for antennas without polarization
squint, the off-diagonal terms will remain significant, even though
the difference between the parallel-hand terms may be negligible.
Hence, full-Sky Jones matrix treatment is necessary for full Stokes
imaging of most observed fields.  This implies a four-fold increase in
the computing load when compared to imaging when primary beam effects
are neglected.

The deconvolution algorithm described in this paper corrects for
systematic effects due to non-ideal primary beams by modeling the
complex antenna aperture illumination for the two orthogonal
polarizations as a function of parallactic angle and antenna pointing
errors.  The antenna aperture functions are used to construct precise
forward and approximate inverse transforms, exploiting the property
that the Sky Jones matrix is approximately unitary.  We have applied
this algorithm to VLA 1.4~GHz imaging and show that the instrumental
Stokes-V is eliminated to an accuracy of better than 10\%, possibly
limited by uncertainties in our model of the primary beam as well as
pointing errors that were not corrected in this reduction because our
algorithm does not yet handle pointing errors and selfcalibration
simultaneously.  Simulations of single pointing observations at
1.4~GHz with the EVLA with typical time-varying antenna pointing
errors show that antenna pointing errors limit the imaging dynamic
range at a level of $\sim 10^5:1$ away from strong sources in a
typical field (imaging dynamic range could be even lower if you are
unlucky, like for the IC2233 case).  Using this algorithm on this
simulated data we demonstrate that the effects of antenna pointing
errors can also be corrected during deconvolution.  This approach can
therefore be used for full-beam full-Stokes imaging.

Finally, we note that in the presence of image plane errors, imaging
and calibration algorithms are more tightly coupled compared to
those appropriate to direction-independent calibration.  Solvers for
parameters which describe direction dependent errors require the
forward transform used during imaging.  Correction for direction
dependent effects is done during image deconvolution and one cannot
produce corrected visibilities independent of full image
deconvolution.  With the advent of higher sensitivity arrays where
many direction dependent errors will need to be accounted for, modern
imaging and calibration software must be designed to easily
accommodate these cases.

%__________________________________________________________________

\begin{acknowledgements}
  All of this work was done using the $\mathcal{CASA}$ package.  We
  thank W.~Brisken for his help in modeling the VLA antenna aperture
  function and M.~A.~Voronkov for his many useful comments.  We thank
  Lynn Matthews for her excellent bandpass calibration of the IC2233
  data.  We have benefited from discussions with Bill Cotton, Rick
  Fisher, Rick Perley and Ken Sowinski.
  
\end{acknowledgements}

\bibliographystyle{aa}
%\begin{thebibliography}{aa}
\bibliography{pbwp_aa}
%\end{thebibliography}

\end{document}